\documentclass{llncs}

\usepackage[british]{babel}
\usepackage{graphicx}
\usepackage{url}
\usepackage[noadjust]{cite}
\usepackage[font=small,labelfont=bf,labelsep=space]{caption}
\begin{document}

\title{Privacy Salience: Taxonomies \\and Research Opportunities}
\titlerunning{Privacy Salience: Taxonomies and Research Opportunities}

\author{Meredydd Williams*, Jason R. C. Nurse and Sadie Creese}
\authorrunning{Williams et al.}

\institute{Department of Computer Science,\newline University of Oxford, Oxford, UK\\
\email{*meredydd.williams@cs.ox.ac.uk}}

\maketitle

\begin{abstract}
Privacy is a well-understood concept in the physical world, with us all desiring some escape from the public gaze. However, while individuals might recognise locking doors as protecting privacy, they have difficulty practising equivalent actions online. Privacy salience considers the tangibility of this important principle; one which is often obscured in digital environments. Through extensively surveying a range of studies, we construct the first taxonomies of privacy salience. After coding articles and identifying commonalities, we categorise works by their methodologies, platforms and underlying themes. While web browsing appears to be frequently analysed, the Internet-of-Things has received little attention. Through our use of category tuples and frequency matrices, we then explore those research opportunities which might have been overlooked. These include studies of targeted advertising and its affect on salience in social networks. It is through refining our understanding of this important topic that we can better highlight the subject of privacy.

\end{abstract} 

\section{Introduction}
\label{sec:one}

Privacy is a well-understood concept in the physical world. We all need some respite from the public gaze to enjoy our lives; indeed, it is essential to natural human development \cite{Berscheid1977}. However, whereas individuals might consider a locked door as protecting one's privacy, they have difficulty practising equivalent actions online \cite{Creese2009}. This can create a number of risks as users might be unaware of the digital dangers they face. Combining the definition of `salience' \cite{OxfordEnglishDictionary2016} with informational privacy \cite{Clarke1999}, we define `privacy salience' as whether ``\textit{informational privacy is prominent in a person's awareness or their memory of past experience}''. This differs slightly from `privacy awareness', which we take to reflect long-term awareness of privacy, such as that which can be improved through educational campaigns. Risk in cyberspace is often intangible \cite{Jackson2005} and research \cite{Adjerid2014} suggests reduced salience can lead to unwise decisions. Some \cite{Williams2016a} have claimed this intangibility could even contribute to the `Privacy Paradox' \cite{Barnes2006}, the disparity between what individuals claim about privacy and how they act. As technology permeates our society and we begin to live our lives `online', privacy salience gains critical importance.

Previous research has considered the topic from a number of angles. For example, John et al. \cite{John2009} conducted several field experiments: two seeking to highlight privacy and one looking to hide the issue. They found that when privacy concerns were primed, participants were less likely to disclose their data. In contrast, Tsai et al. \cite{Tsai2009} modified search engine interfaces to promote privacy-respecting results. Their analysis of 15,000 queries found that their alterations encouraged prudent selections. Adjerid et al. \cite{Adjerid2013} studied how the provision of privacy information could influence user actions. They discovered that a delay of only 15 seconds between notice and decision could lead to less-private behaviour. Although previous studies concern a range of platforms and themes, the field has seen little systemisation of knowledge. Neither an extensive literature review nor a taxonomy have yet been produced: instruments which can both structure existing work and highlight future opportunities. Such developments are crucial to ensure that new studies do not overlook the varied findings of past research.

Therefore, we develop three extensive taxonomies of privacy salience literature, classifying studies by the themes they concern, the methodologies they apply and the platforms on which they are based. We select these factors as we believe they best encapsulate the content of the articles. Through a data-driven process of inductive coding \cite{Thomas2006}, we formulate categories ranging from social networks to smartphones, privacy seals to permissions. We classify our surveyed articles within these groups and discuss the literature most relevant to each section. We move on to investigate those category combinations, whether (\textit{Methodology},\textit{Platform}), (\textit{Methodology},\textit{Theme}) or (\textit{Platform},\textit{Theme}), which are both feasible and underexplored. Our frequency matrices, populated by this series of category tuples, enable identification of both popular research areas and potential lacunas. For example, while privacy documents were often studied during web browsing, salience is rarely explored in the Internet-of-Things (IoT). This is of concern as unfamiliar devices could potentially mask the topic of privacy. We conclude by recommending both future work and potential extensions to our taxonomies.

The remainder of our paper is structured as follows. Section \ref{sec:two} discusses our methodology, including literature selection, exclusion criteria, coding processes and taxonomy construction. Section \ref{sec:three} then explores our three taxonomies in detail, highlighting relevant previous literature. In Section \ref{sec:four} we present our distribution matrices and identify opportunities for future research. Finally, we conclude the paper in Section \ref{sec:five} and reflect on possible extensions to this work.

\newpage

\section{Methodology}
\label{sec:two}

We first outline our definitions, before describing our processes to select and exclude existing work. We continue by discussing our inductive coding processes \cite{Thomas2006} and how our privacy salience taxonomies were constructed.

\vspace{-0.5em}

\subsection*{Definitions}

To ensure our taxonomies are representative of the literature, we should precisely define our terms. The Oxford English Dictionary defines salience to be ``\textit{[t]he quality or fact of being more prominent in a person's awareness or in his memory of past experience}'' \cite{OxfordEnglishDictionary2016}. As privacy can be a nebulous topic, we scope our definition to encompass informational privacy. Clarke \cite{Clarke1999} described this concept as ``\textit{the interest an individual has in controlling, or at least significantly influencing, the handling of data about themselves}''. Therefore, to reiterate, we define privacy salience as as whether ``\textit{informational privacy is prominent in a person's awareness or their memory of past experience}''. This differs slightly from `privacy awareness', which we take to reflect long-term awareness of privacy, such as that which can be improved through educational campaigns.

We also explicitly specify what we consider to be a taxonomy. The Oxford English Dictionary \cite{OxfordEnglishDictionary2016a} defines a taxonomy as a ``\textit{particular system of classification}'', and in this work we classify privacy salience research. De Hoog \cite{DeHoog1981} explains how construction consists of three parts: ordering, representation and nomenclature. In terms of ordering, categories should be arranged in a certain order and this order expressed through ``\textit{character correlation}''. For representation, the elements should be ``\textit{maximally simple}'' and atomic in character. Finally, the categories should be named formally to ensure the structure is usable. We incorporated these key principles into the development of our taxonomies.

\vspace{-0.5em}

\subsection*{Literature Selection}

We first surveyed existing research to identify those works which concerned privacy salience. We did not explicitly constrain ourselves to particular disciplines, as we sought to explore the topic from multiple angles. Accordingly, we conducted our literature search on a wide range of databases from a variety of fields. These consisted of Google Scholar, Scopus, Web of Science, SpringerLink, JSTOR and Mendeley (general); IEEE Xplore, ACM Digital Library, CiteSeerX and DBLP Computer Science (computer science, Human-Computer Interaction (HCI) and cyber security); ScienceDirect (sciences); the Social Science Research Network (SSRN) (social sciences) and HeinOnline (law). These databases index those fields from which privacy salience research frequently originates, such as HCI and psychology. With engines such as Google Scholar searching broader academia and SSRN considering the social sciences, we retrieved work from a wide range of disciplines. Since we frequently located the same articles in multiple search results, we are confident the literature was well surveyed.

We also used a variety of search terms to ensure all works considering privacy salience were identified. We began with the synonymous terms `\textit{privacy salience}' and `\textit{privacy saliency}', in addition to `\textit{privacy tangibility}' due to its similar definition. Being cognisant that the `\textit{privacy salience}' term only gained popularity in the late 2000s, we also searched for `\textit{privacy awareness}'. Frequently salience was not mentioned in articles, even though studies considered the effects of highlighting policies and notices. For this reason, we also used `\textit{privacy policies}', `\textit{privacy seals}', `\textit{privacy notices}', `\textit{privacy warnings}', `\textit{privacy indicators}' and `\textit{privacy nudges}'. By expanding our list of phrases, we successfully identified articles which might have been otherwise overlooked.

To further survey this topic, we undertook literature snowballing through the references and citations of identified works. This was performed in the systematic method of Wohlin \cite{Wohlin2014}, with extensive backwards and forwards snowballing following our database search. This collection was then manually-filtered to verify that selected works concerned the topic. This ensured that we did not sacrifice quantity for quality by expanding our search terms. Although terms can never be fully exhaustive, our broad selection concerns topics frequently associated with privacy salience. Since many search results were sorted by both relevance and citation count, it is unlikely we overlooked articles of significance. 

\vspace{-0.5em}

\subsection*{Exclusion Criteria}

To complement our search term expansion, we strengthened our exclusion criteria. Firstly, we only analysed articles in the English language to ensure works could be judged fairly. Secondly, we verified whether our search results actually concerned privacy salience, of which a majority did not. This is simply an artefact of database searches, where works can refer to `\textit{privacy}' or `\textit{salience}' but not both in combination. Furthermore, through our use of associated terms such as `\textit{privacy policies}', we retrieved many articles which considered these documents but not their effect on salience. These works were filtered out at this stage.

Thirdly, we excluded papers which directly duplicated research. For example, two articles by Hughes-Roberts \cite{Hughes-Roberts2014a,Hughes-Roberts2015a} concerned the development of the same social networking interface. In these cases, we included the most recent paper as would be more likely to possess additional findings. For a similar reason, when multiple databases returned different versions of a work, we selected the most recent instance. Finally, as a means of ensuring our research was of a high quality, we excluded articles which were not peer-reviewed. Although this approach might have reduced the breadth of our survey, it is important that taxonomies are constructed on credible works.

\vspace{-1em}

\subsection*{Coding Process}

Inductive reasoning can be beneficial when conducting research which has not been previously attempted. Since we are the first to either survey privacy salience or construct taxonomies on the topic, inductive coding appeared most appropriate. We conformed to the popular approach of Thomas \cite{Thomas2006} which consists of data cleaning, text analysis, category creation, overlapping coding and category refinement. We defined our coding units physically \cite{Stemler2001}, based on the natural boundaries of each article. We began coding by ensuring all text was legible, accessible and downloaded in a persistent format. We then analysed the documents to familiarise ourselves with the main topics. 

After studying our articles on multiple occasions, we created initial groups based on their main concepts. As inductive coding progressed, we recognised that our categories clustered around methodologies, platforms and themes. We believe these factors best encapsulate the paper content, since they concern the research technique, the target of research and the research topic. Methodologies can influence how an issue is approached: while literature reviews reflect on an issue, field experiments conduct empirical studies. Although article frequency does not directly indicate which techniques are most fruitful, it acts as a useful proxy. While certain methodologies might be more appropriate for privacy salience research than others, we still expect a general trend between frequency of use and utility. We define platform as the domain on which salience is analysed, with instances ranging from social networks \cite{Stutzman2010} to smartphones \cite{Rajivan2016}. As privacy is inherently contextual, findings on one platform might differ from those on others. Although each article concerns privacy salience, the topic is explored through a range of themes. For example, while some researchers study the effect of privacy policies \cite{Vail2008}, others analyse the influence of framing \cite{Johnson2002}. 

It was at this stage we decided to develop three distinct taxonomies rather than a combined chart. While our factors each offered interesting insights, they differed excessively for a single structure. Although we considered a taxonomy with methodology, platform and theme top-level categories, this introduced unnecessary complexity. By developing three separate taxonomies and classifying our works, we can identify those combinations which might be underexplored. 

Thomas' fourth coding procedure \cite{Thomas2006} accepts that texts can be coded into multiple categories, or indeed no categories at all. This is useful in the case of themes, as a study might concern a number of topics. For example, Yang et al. \cite{Yang2015} investigated the influence of both privacy policies and trust seals, and their work should not be excluded from either group. As articles predominantly used one methodology and one platform, these factors support single categories. Finally, we refined our groups to ensure clarity and consistency. For example, although one work might analyse social media disclosures \cite{Hughes-Roberts2015a} and another might study Facebook behaviour \cite{Stutzman2010}, both concern social networks. This was crucial for de Hoog's taxonomy `ordering' principle \cite{DeHoog1981}, as we ensured similar elements were grouped consistently.

\vspace{-1em}

\subsection*{Taxonomy Completion}

We continued our construction processes in compliance with the `representation' and `nomenclature' principles \cite{DeHoog1981}. Where categories contained otherwise differing elements, further subdivisions were made. For example, although policies and seals both highlight privacy, their approaches are far from identical. We next created representative names for our categories, complying with the `nomenclature' principle \cite{DeHoog1981}. Naming was undertaken iteratively, refining definitions as categories evolved. Titles aimed to encapsulate commonalities in a group; for example, privacy policies and notices could both be considered types of privacy document.

\section{Privacy Salience Taxonomies}
\label{sec:three}

We begin by outlining category metrics before discussing our three taxonomies in detail. Through our database search and snowballing process, we received over 1000 potential results. After manual filtering, we found only 76 articles actually concerned privacy salience, with the other works just matching on search terms. This list was further reduced to 73 papers based on our aforementioned exclusion criteria. While our search dates were not constrained, our selected literature ranged from 1977 to June 2016. Research was conducted through a range of methodologies, with Field Experiments and Tool Development appearing most prevalent. In terms of platform, the Web was found to be most popular, with Social Networks also frequently explored. Interfaces were the most commonly-identified theme, followed by Framing and privacy Documents. Table \ref{fig:table} below presents our categories and the quantitative distribution of works. As articles could concern multiple topics, this is reflected in our larger theme totals.

\vspace{-0.5em}

\begin{table}[ht]
  \centering
    \includegraphics[width=1.0\textwidth]{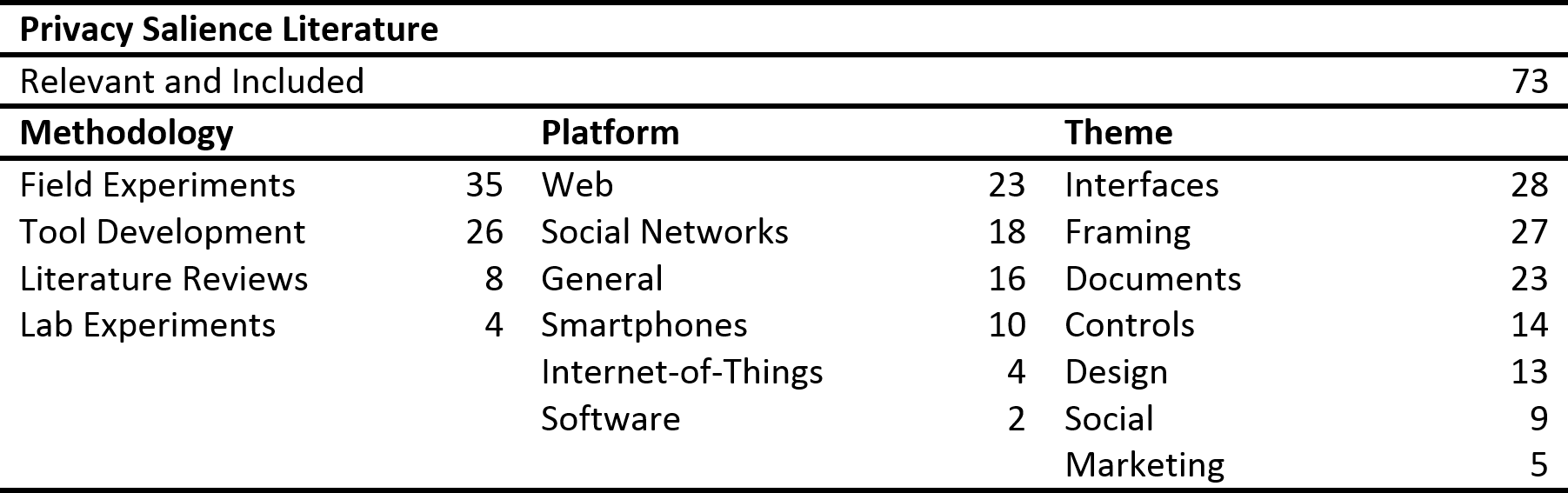}
    \setlength{\abovecaptionskip}{2pt}
    \caption{Literature distribution}
    \label{fig:table}
\end{table}

\vspace{-4em}

\subsection*{Methodology Taxonomy}

As shown below in Figure \ref{fig:methodologytaxonomy}, we subdivided methodologies into four approaches: Literature Reviews, Lab Experiments, Field Experiments and Tool Development. Methodologies were identified by comparing research techniques with standard definitions. Although a minority of articles possessed multiple approaches, we classified based on the predominant methodology. For example, when an application is created and then evaluated through a field study, it would be categorised within the Tool Development group \cite{LaRose2007}. Although we considered supporting multiple methodologies, this approach would have introduced complexity not warranted by our literature. 

\begin{figure}[ht]
  \centering
    \includegraphics[width=1.0\textwidth]{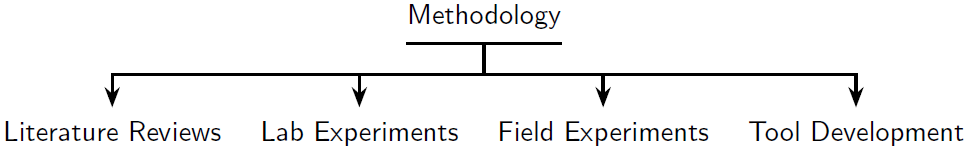}
    \caption{Methodology taxonomy}
    \label{fig:methodologytaxonomy}
\end{figure}

\textbf{Literature Reviews} study an existing body of work to derive novel findings. For example, Aguirre et al. \cite{Aguirre2016} drew on prior research to discuss how firms can best manage consumer relationships. They found that although service personalisation offers benefits, it can increase the salience of privacy risk. Cichy and Salge \cite{Cichy2015} also analysed previous work, studying 35 years of privacy discourse in The New York Times. Through considering social norms and topic salience, they found perceptions to be susceptible to myopia and manipulation.

\textbf{Lab Experiments} are empirical studies conducted in well-controlled environments. For example, a 24-person study was used to evaluate three privacy-enhancing extensions \cite{Schaub2016}. The researchers found that although the plug-ins highlighted data collection, concerns were mitigated by the applications themselves. Vemou et al. \cite{Vemou2014} established social network accounts to analyse profile registration and privacy policies. After exploring third-party access and audience management, they concluded that simplified settings might improve salience.

\textbf{Field Experiments} are undertaken in realistic environments, benefiting from greater ecological validity than lab studies. To explore the effects of framing, 280 participants were tasked to create a social networking profile \cite{Adjerid2013}. It was found that when privacy notices were followed by time delays, data disclosure increased. John et al. \cite{John2009} conducted three user studies, with the former increasing privacy salience and the latter two disguising the topic. They saw that even when risks are low, people refuse to disclose when their concerns are primed.

\textbf{Tool Development} concerns research which develops interfaces or applications to increase privacy salience. For example, PrivAware was a social networking tool which highlighted information loss \cite{Becker2009}. The system could infer personal details with 60\% accuracy and gave recommendations for friend deletion. Lipford et al. \cite{Lipford2008} developed an `audience view', allowing users to observe their profiles as others do. After adding this tab to the Facebook interface, they found individuals better-understood the consequences of their actions.

\vspace{-0.5em}

\subsection*{Platform Taxonomy}

Platforms were defined based on the domain in which privacy salience research was undertaken. As presented below in Figure \ref{fig:platformtaxonomy}, we distinguished between six categories: General, Web, Social Networks, Mobile, Software and Internet-of-Things. We found the General class to be beneficial, as several works \cite{Acquisti2009,Johnson2002,Frey1986} consider salience without reference to a particular platform. As Social Networks are distinctive portals which may be accessed via either web browsers or smartphones, we deemed these to define a separate category.

\begin{figure}[ht]
  \centering
    \includegraphics[width=1.0\textwidth]{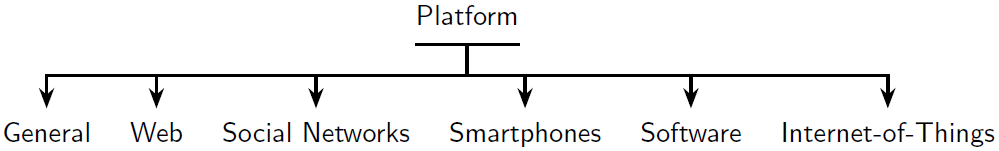}
    \caption{Platform taxonomy}
    \label{fig:platformtaxonomy}
\end{figure}

\textbf{General} works study privacy salience without direct consideration of a specific environment. For example, Acquisti \cite{Acquisti2009} discussed the importance of `nudging' for realigning user behaviour. He emphasised the benefits of soft paternalism, but did not constrain privacy concepts to a particular domain. In a 1979 study, Reamer \cite{Reamer1979} analysed whether guarantees of confidentiality paradoxically reduced disclosure. He found survey participants which were assured anonymity were less likely to respond, suggesting salience had an effect.

The \textbf{Web} category concerns works which study privacy salience during web browsing. Tsai et al. \cite{Tsai2009} found that privacy indicators on search engines can encourage prudent behaviour. Through their use of the Privacy Finder tool, they found sites were more popular if annotated as privacy-respecting. Plug-ins can also analyse online behaviour, such as the Privacy Fox browser extension \cite{Arshad2004}. This application both translated policies into short notices and highlighted website practices which might cause concern.

Since \textbf{Social Networks} support the interaction of online individuals, they are of great interest to privacy researchers. During a 6-week trial, Facebook users were nudged to remember their post audiences \cite{Wang2014}. By illustrating the potential consequences of their actions, unintended disclosures were reduced. Bonneau and Preibusch \cite{Bonneau2010} evaluated 45 sites in a comprehensive analysis of social network protections. They saw that since data disclosure can be reduced by salient privacy, this influences interface design.

The \textbf{Smartphones} category concerns those works which study mobile phones and their apps. For example, the AppOps tool was used to highlight the data shared between smartphone applications \cite{Almuhimedi2015}. When the consequences of lax privacy were illustrated, over half the participants changed their permissions. Balebako et al. \cite{Balebako2015} explored the timing of privacy notices through an Android field experiment. They found salience was increased more by in-app dialogs than those shown before installation.

The \textbf{Software} section concerns works which analyse desktop applications, rather than online portals or mobile apps. In a similar manner to Balebako et al. \cite{Balebako2015}, a 222-person study explored how notice timing affects user behaviour \cite{Good2007}. The researchers found that risky installations were reduced by summarising license agreements. Bravo-Lillo et al. \cite{Bravo-Lillo2013} modified user interfaces to highlight security and privacy threats. In their study of dialog messages, they discovered salient warnings could reduce dangerous installations.

The \textbf{Internet-of-Things} category concerns the privacy analyses of smart devices. For example, it was proposed that RFID privacy salience could be increased by personal privacy assistants \cite{Konomi2004}. This gadget would process tag data and display risks on a mobile interface. Gisch et al. \cite{Gisch2007} developed the Privacy Badge, an awareness tool specifically designed for small devices. Their data loss visualisations were evaluated through a user study, which found their application to be usable and informative.

\subsection*{Theme Taxonomy}

Themes refer to the range of topics considered in the privacy salience literature. We identified 17 themes through inductive coding, ranging from Nudging \cite{Acquisti2009} to Adverts \cite{Wang2015a} to privacy Settings \cite{Stutzman2010}. Through our process of taxonomy construction, these classes were grouped under 7 top-level categories. This structure is presented below in Figure \ref{fig:themetaxonomy}. This design both highlights theme commonalities and assists the lacuna identification discussed in Section \ref{sec:four}. While we considered limiting works to a single theme, such an approach would underrepresent the secondary topics found in articles. Recognising that the vast majority of papers did concern several topics, we decided each work could support multiple themes.

\vspace{-1.0em}

\begin{figure}[ht]
  \centering
    \includegraphics[width=1.0\textwidth]{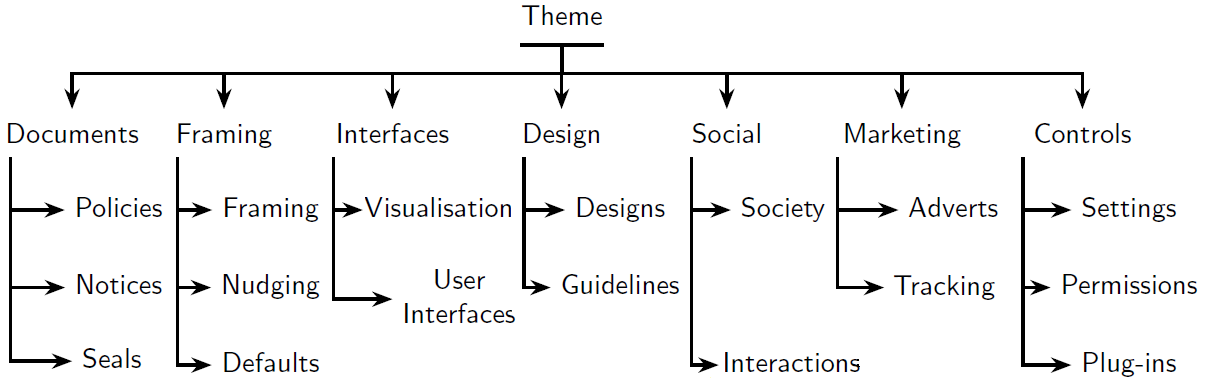}
    \caption{Theme taxonomy}
    \label{fig:themetaxonomy}
\end{figure}

\vspace{-0.8em}

\textbf{Documents} refer to a range of privacy statements, such as policies, notices and seals. In one study, documents were presented which either concerned tech company protections or their negative activities \cite{Marreiros2016}. The authors found disclosure decreased even when they highlighted positive behaviour, suggesting privacy salience influenced action. Hui et al. \cite{Hui2007} analysed how privacy statements and TRUSTe seals affect user behaviour. They discovered that seals had little impact, suggesting increased salience does not always translate into action.

\textbf{Framing} relates to the way in which information or choices are presented. For example, two online surveys explored whether default responses affect privacy decisions \cite{Johnson2002}. When choices were made opt-in, the researchers found that agreement increased by 30\%. Joinson et al. \cite{Joinson2008} discovered that `prefer not to say' options were used more frequently when privacy was salient. After priming the topic through a privacy questionnaire, they also found that sensitive questions were commonly avoided.

The \textbf{Interfaces} category refers to both visualisations and user interface developments. Kani-Zabihi and Helmhout \cite{Kani-Zabihi2012} discussed online interactive privacy features; tools designed to support user decision-making. They described an enquiry system which helps individuals discuss their privacy concerns with service providers. In another study, graphical warnings were evaluated on mobile interfaces \cite{Christin2013}. The researchers discovered that after their dialogs were presented, 70\% of the participants claimed they would change their permissions.

\textbf{Design} guidelines can inform the development of technologies which make privacy salient. One article \cite{Lederer2004} presented five pitfalls for privacy design, including obscuring information flow and inhibiting existing practice. The authors explained how ``\textit{users can make informed use of a system only when they understand the scope of its privacy implications}''. Schaub et al. \cite{Schaub2015} outlined requirements and best practices for privacy notice design. They went on to discuss how dialog messages could be challenged by Internet-of-Things interfaces.

Privacy is a \textbf{Social} construct and therefore many studies concern user interactions. For example, eight Facebook users were interviewed on the topic of online friendship \cite{Houghton2010}. The researchers discovered that unwise disclosures often originated from a lack of privacy salience. Ziegeldorf et al. \cite{Ziegeldorf2015} observed that comparisons are a natural behaviour, with individuals evaluating their actions against those of their peers. Their nudging system aimed to incentivise privacy by highlighting the behaviour of others.

With \textbf{Marketing} frequently raising privacy concerns, this category concerns adverts and tracking. A study of 447 users analysed how smartphone ad awareness affects privacy perceptions \cite{Wang2015a}. Individuals were found to make better privacy decisions when informed of data use procedures. Goldfarb and Tucker \cite{Goldfarb2011} also analysed marketing, studying how relevance and obtrusiveness influence purchasing intent. They discovered that while ads are effective when either salient or targeted, combined approaches trigger privacy concerns.

\textbf{Controls} refer to the permissions and configurations which affect user privacy. 444 students were surveyed to gauge the popularity of `friends-only' Facebook settings \cite{Stutzman2010}. Although respondents were not aware of their exposure, privacy discussions were found to increase salience. Malandrino et al. \cite{Malandrino2013} analysed data leakage through a browser plug-in. By highlighting the information collected by third-party services, they increased the salience of privacy violations.

\section{Research Gaps and Opportunities}
\label{sec:four}

By highlighting those areas not frequently explored, we sought to identify potential research opportunities. We first considered the prevalence of our methodologies, platforms and themes. Although our surveyed works concern a wide variety of themes, we constrained our analyses to the top-level categories. While this simplification sacrificed a degree of depth, it assisted the identification of sparse research areas.  

We next considered all possible 2-tuples for (\textit{Methodology},\textit{Platform}), (\textit{Methodology},\textit{Theme}) and (\textit{Platform},\textit{Theme}) combinations. Again, use of all 17 low-level themes would have introduced significant complexity to this process. Furthermore, as the vast majority of combinations would not have been previously explored, we would have no direction in selecting future opportunities. While we considered the use of triples, this approach challenges tabular visualisation (as explained shortly). By analysing our 2-tuples individually, we could identify combinations with greater flexibility. 

\begin{figure}[ht]
  \centering
    \includegraphics[width=1.0\textwidth]{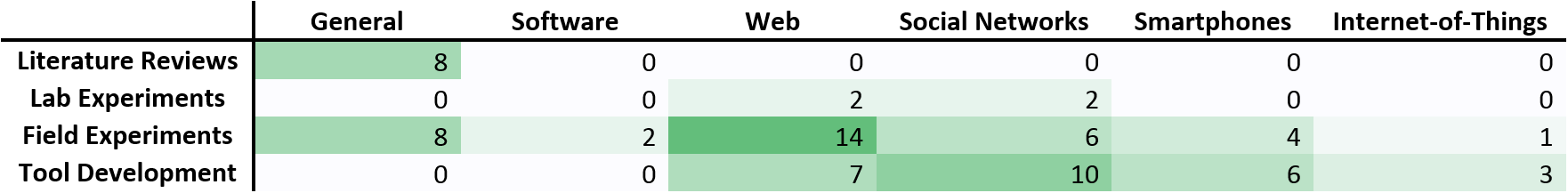}
    \caption{(\textit{Methodology},\textit{Platform}) frequency matrix}
    \label{fig:methplat}
\end{figure}

\begin{figure}[ht]
  \centering
    \includegraphics[width=1.0\textwidth]{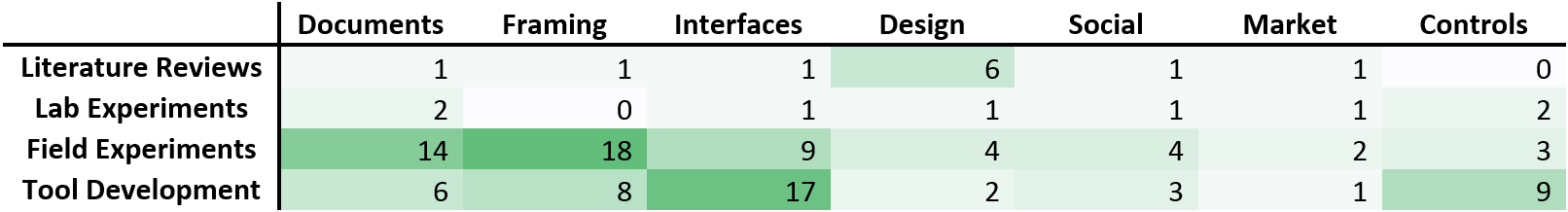}
    \caption{(\textit{Methodology},\textit{Theme}) frequency matrix}
    \label{fig:meththeme}
\end{figure}

\begin{figure}[ht]
  \centering
    \includegraphics[width=1.0\textwidth]{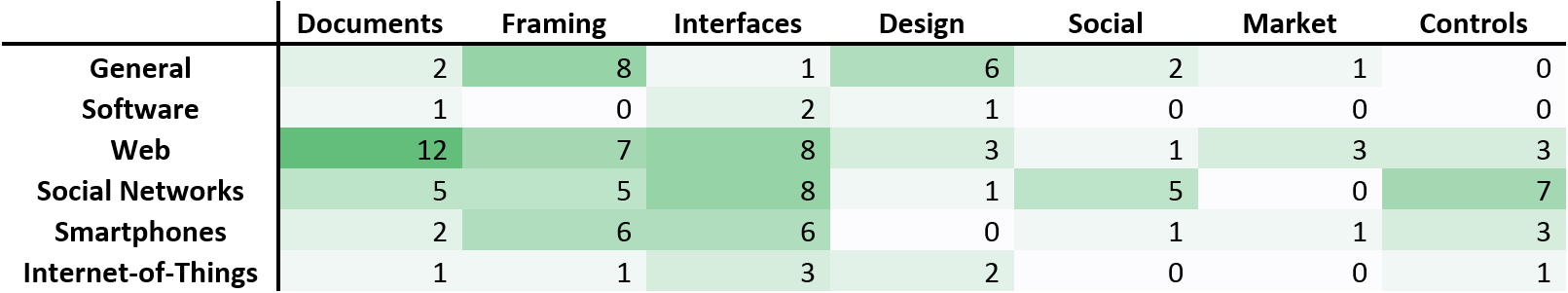}
    \caption{(\textit{Platform},\textit{Theme}) frequency matrix}
    \label{fig:plattheme}
\end{figure}

Through analysing the tuple frequency in our literature, we populated the entries in three matrices. Figure \ref{fig:methplat} presents (\textit{Methodology},\textit{Platform}) works, Figure \ref{fig:meththeme} concerns (\textit{Methodology},\textit{Theme}) and Figure \ref{fig:plattheme} relates to (\textit{Platform},\textit{Theme}) combinations. As we supported one methodology and platform per paper (due to the predominance of this distribution), the first matrix was simple to complete. Since articles could concern several themes, each work could conform to multiple (\textit{Methodology},\textit{Theme}) and (\textit{Platform},\textit{Theme}) tuples. While this could skew the frequency of non-theme categories, it assisted the identification of less-populated research areas. 

By comparing the frequency of matrix entries, we discerned which combinations are most popular. For example, Web Field Experiments were prevalent (14 instances), as were Tools with privacy Interfaces (17). Privacy Documents were commonly analysed during Web browsing (12), reflecting the research interest in policies and seals. Finally, these tuple frequencies were colour-coded based on their magnitude. This assisted identification of those areas least investigated in existing work. Although many combinations have been infrequently explored, we continue by exploring those sparse intersections with the greatest viability. A low frequency alone does not imply an opportunity, as some combinations are less feasible than others. However, whereas Social interactions might not be best analysed through Lab Experiments, several combinations do appear viable.

\vspace{-2em}

\vspace{-2em}

\subsection*{Research Opportunities}

Through inspection of our frequency matrices, we observed three main areas with feasible opportunities. This is not to say that other alternatives are not valuable, such as much-needed Literature Reviews and Software studies, but the below topics appear of particular interest.

Due to the novelty of the platform, the Internet-of-Things (IoT) appears underexplored. As presented in Figure \ref{fig:methplat}, few Lab or Field Experiments have analysed IoT salience. Although privacy risks can obscured by our current technologies, unfamiliar devices might exacerbate this issue. Researchers could use controlled lab studies to investigate how much data is leaked by novel products. This could be complemented by field experiments exploring how these tools are actually used. We could analyse how salient privacy is in smart home environments; locations which are personal and increasingly popular. Figure \ref{fig:plattheme} suggests that while Social Network Controls are frequently studied, several platforms do not receive similar attention. IoT settings might be unfamiliar, hidden or challenging to adjust, resulting in unintentional data disclosure. Future work could explore whether privacy salience can enhanced by simplifying these interfaces.

Figure \ref{fig:meththeme}, the (\textit{Methodology},\textit{Theme}) matrix, suggests that while Tool Developments are popular, principles of their Design are rarely considered. While guidelines have been constructed for Privacy Enhancing Technologies (PETs) \cite{Federrath2001}, the topic of salience has not been approached. Social network modifications \cite{Hughes-Roberts2015a} and smartphone interfaces \cite{Almuhimedi2015} have successfully highlighted privacy, resulting in improved user behaviour. For these achievements to be replicated generally, researchers should develop design principles for what could be considered `Salience-Enhancing Technologies'. 

Much of the web is supported by affiliate networks and targeted advertising. This has led many individuals to decry the increasingly-personalised nature of online ads. Despite this fact, Figure \ref{fig:plattheme} suggests that few have explored the relationship between Marketing and privacy salience. Companies face the challenge of advertising their services without priming concerns. Future research could study whether online portals can respect user privacy while delivering targeted advertising. Alternatively, with the increasing popularity of ad-blocking software, researchers could explore how salience is affected when adverts are removed. It is of particular surprise these Marketing studies have not analysed Social Networks, especially considering Facebook's advertising strength. Researchers could investigate how privacy concerns are suppressed and suggest approaches for highlighting these ads.

\newpage

\section{Conclusions}
\label{sec:five}

In this work we have considered at length the important topic of privacy salience. We begun by surveying a wide range of existing literature before identifying common themes and topics. Through iteratively refining these categories we developed the first taxonomies on privacy salience. We structured these divisions on methodologies, platforms and themes; factors which encapsulate the content of existing research. By classifying prior work within these taxonomies, we systematise the knowledge in the literature. We proceeded by analysing category frequency and which factor combinations are most popular. We found that field studies of the web were prevalent, as were tools with privacy interfaces. Finally, through exploring our colour-coded matrices, we identified opportunities for future research. These include IoT field experiments and investigations of online marketing. We also recommend further analyses of privacy settings, whether in ubiquitous environments or desktop computers.

While we believe our taxonomies support the study of privacy salience, we accept several limitations to our work. A minority of articles possessed multiple methodologies or studied a variety of platforms. Due to the sparsity of these instances, we constrained our analyses in the interest of simplicity. An expanded future work would reflect the diversity inherent in many privacy studies. While we chose methodologies, platforms and themes to structure our taxonomies, we accept that there are several alternatives. We could have divided research by discipline as a means of highlighting interesting fields. Although this was considered, we felt it would be challenging to divide disciplines in an objective fashion. Alternatively, we could have categorised articles based on the metrics they analysed. For example, while many works studied disclosure \cite{Reamer1979,John2009}, others explored settings alteration \cite{Almuhimedi2015,Christin2013}. In future analyses, metrics could be incorporated to consider how salience can be best investigated.

While our research presents an initial analysis, we envisage many instances of future work. As a means of validating our taxonomies, we could invite privacy experts to categorise the literature. In this approach, we  would solicit a panel to analyse and classify each of our surveyed works. Through exploring their range of classifications, we could refine the consistency of our structures. Although research has been conducted in a range of platforms, there have been few comparative studies. Whereas privacy might be salient on a familiar web browser, the topic could be obscured by smartphone interfaces. Through conducting user studies on multiple platforms, we could explore how contexts affect privacy concerns. Finally, metrics should be studied to enable us to deconstruct the concept of privacy salience. While disclosure might relate to notions of confidentiality, privacy settings might be concerned with user control. As novel technologies proliferate, we believe privacy salience will only become increasingly important.

\newpage

\bibliographystyle{splncs03}
\bibliography{bib}

\begin{thebibliography}{10}
\providecommand{\url}[1]{\texttt{#1}}
\providecommand{\urlprefix}{URL }

\bibitem{Acquisti2009}
Acquisti, A.: {Nudging privacy: The behavioral economics of personal
  information}. IEEE Security {\&} Privacy  7(6),  82--85 (2009)

\bibitem{Adjerid2014}
Adjerid, I., Acquisti, A., Loewenstein, G.: {Framing and the malleability of
  privacy choices}. In: Proceedings of the 13th Workshop on the Economics of
  Information Security (2014)

\bibitem{Adjerid2013}
Adjerid, I., Acquisti, A., Brandimarte, L., Loewenstein, G.: {Sleights of
  privacy: Framing, disclosures, and the limits of transparency}. In:
  Proceedings of the Ninth Symposium on Usable Privacy and Security (2013)

\bibitem{Aguirre2016}
Aguirre, E., Roggeveen, A., Grewal, D., Wetzels, M.: {The
  personalization–-privacy paradox: Implications for new media}. Journal of
  Consumer Marketing  33(2) (2016)

\bibitem{Almuhimedi2015}
Almuhimedi, H., Schaub, F., Sadeh, N., Adjerid, I., Acquisti, A., Gluck, J.,
  Cranor, L.F., Agarwal, Y.: {Your location has been shared 5,398 times!: A
  field study on mobile app privacy nudging}. In: Proceedings of the 33rd
  Annual ACM Conference on Human Factors in Computing Systems. pp. 787--796
  (2015)

\bibitem{Arshad2004}
Arshad, F.: {Privacy fox - A JavaScript-based P3P agent for Mozilla Firefox}.
  Privacy Policy, Law, and Technology  17,  801--810 (2004)

\bibitem{Balebako2015}
Balebako, R., Schaub, F., Adjerid, I., Acquisti, A., Cranor, L.: {The impact of
  timing on the salience of smartphone app privacy notices}. In: Proceedings of
  the 5th Annual ACM CCS Workshop on Security and Privacy in Smartphones and
  Mobile Devices. pp. 63--74 (2015)

\bibitem{Barnes2006}
Barnes, S.: {A privacy paradox: Social networking in the United States}. First
  Monday  11(9) (2006)

\bibitem{Becker2009}
Becker, J., Chen, H.: {Measuring privacy risk in online social networks}. In:
  Proceedings of the 2009 Web 2.0 Security and Privacy Workshop (2009)

\bibitem{Berscheid1977}
Berscheid, E.: {Privacy: A hidden variable in experimental social psychology}.
  Journal of Social Issues  33(3),  85--101 (1977)

\bibitem{Bonneau2010}
Bonneau, J., Preibusch, S.: {The privacy jungle: On the market for data
  protection in social networks}. Economics of Information Security and Privacy
   (2010)

\bibitem{Bravo-Lillo2013}
Bravo-Lillo, C., Komanduri, S., Cranor, L., Reeder, R., Sleeper, M., Downs, J.,
  Schechter, S.: {Your attention please: Designing security-decision UIs to
  make genuine risks harder to ignore}. In: Proceedings of the Ninth Symposium
  on Usable Privacy and Security (2013)

\bibitem{Christin2013}
Christin, D., Michalak, M., Hollick, M.: {Raising user awareness about privacy
  threats in participatory sensing applications through graphical warnings}.
  In: Proceedings of International Conference on Advances in Mobile Computing
  {\&} Multimedia. pp. 445--455 (2013)

\bibitem{Cichy2015}
Cichy, P., Salge, T.: {The evolution of privacy norms: Mapping 35 years of
  technology-related privacy discourse, 1980-2014}. In: 2015 International
  Conference on Information Systems (2015)

\bibitem{Clarke1999}
Clarke, R.: {Introduction to dataveillance and information privacy, and
  definitions of terms}. Tech. rep. (1999),
  \url{http://www.qatar.cmu.edu/iliano/courses/10F-CMU-CS349/slides/privacy.pdf}

\bibitem{Creese2009}
Creese, S., Lamberts, K.: {Can cognitive science help us make information risk
  more tangible online?} In: Proceedings of the WebSci'09 (2009)

\bibitem{Federrath2001}
Federrath, H.: {Designing privacy enhancing technologies}. Springer (2001)

\bibitem{Frey1986}
Frey, J.: {An experiment with a confidentiality reminder in a telephone
  survey}. Public Opinion Quarterly  50(2),  267--269 (1986)

\bibitem{Gisch2007}
Gisch, M., {De Luca}, A., Blanchebarbe, M.: {The privacy badge: A
  privacy-awareness user interface for small devices}. In: Proceedings of the
  4th International Conference on Mobile Technology, Applications, and Systems.
  pp. 583--586 (2007)

\bibitem{Goldfarb2011}
Goldfarb, A., Tucker, C.: {Online display advertising: Targeting and
  obtrusiveness}. Marketing Science  30(3),  389--404 (2011)

\bibitem{Good2007}
Good, N., Grossklags, J., Mulligan, D., Konstan, J.: {Noticing notice: A
  large-scale experiment on the timing of software license agreements}. In:
  Proceedings of the SIGCHI Conference on Human Factors in Computing Systems.
  pp. 607--616 (2007)

\bibitem{DeHoog1981}
de~Hoog, G.: {Methodology of taxonomy}. Taxon  30(4),  779--783 (1981)

\bibitem{Houghton2010}
Houghton, D., Joinson, A.: {Privacy, social network sites, and social
  relations}. Journal of Technology in Human Services  28(1-2),  74--94 (2010)

\bibitem{Hughes-Roberts2014a}
Hughes-Roberts, T., Kani-Zabihi, E.: {On-line privacy behavior: Using user
  interfaces for salient factors}. Journal of Computer and Communications
  2(4),  220 (2014)

\bibitem{Hughes-Roberts2015a}
Hughes-Roberts, T.: {Reminding users of their privacy at the point of
  interaction: The effect of privacy salience on disclosure behaviour}. Human
  Aspects of Information Security, Privacy, and Trust in Lecture Notes in
  Computer Science  9190,  347--356 (2015)

\bibitem{Hui2007}
Hui, K., Teo, H., Lee, S.: {The value of privacy assurance: An exploratory
  field experiment}. MIS Quarterly  31(1),  19--33 (2007)

\bibitem{Jackson2005}
Jackson, J., Allum, N., Gaskell, G.: {Perceptions of risk in cyberspace}.
  Edward Elgar (2005)

\bibitem{John2009}
John, L., Acquisti, A., Loewenstein, G.: {The best of strangers: Context
  dependent willingness to divulge personal information}. In: The Best of
  Strangers. Pittsburgh, Carnegie Mellon-University (2009)

\bibitem{Johnson2002}
Johnson, E., Bellman, S., Lohse, G.: {Defaults, framing and privacy: Why opting
  in-opting out}. Marketing Letters  13(1),  5--15 (2002)

\bibitem{Joinson2008}
Joinson, A., Paine, C., Buchanan, T., Reips, U.: {Measuring self-disclosure
  online: Blurring and non-response to sensitive items in web-based surveys}.
  Computers in Human Behavior  24(5),  2158--2171 (2008)

\bibitem{Kani-Zabihi2012}
Kani-Zabihi, E., Helmhout, M.: {Increasing service users' privacy awareness by
  introducing on-line interactive privacy features}. Information Security
  Technology for Applications in Lecture Notes in Computer Science  7161,
  131--148 (2012)

\bibitem{Konomi2004}
Konomi, S.: {Personal privacy assistants for RFID users}. In: Proceedings of
  the International Workshop Series on RFID 2004. pp. 1--6 (2004)

\bibitem{LaRose2007}
LaRose, R., Rifon, N.: {Promoting i-safety: Effects of privacy warnings and
  privacy seals on risk assessment and online privacy behavior}. Journal of
  Consumer Affairs  41(1),  127--149 (2007)

\bibitem{Lederer2004}
Lederer, S., Hong, J., Dey, A., Landay, J.: {Personal privacy through
  understanding and action: Five pitfalls for designers}. Personal and
  Ubiquitous Computing  8(6),  440--454 (2004)

\bibitem{Lipford2008}
Lipford, H., Besmer, A., Watson, J.: {Understanding privacy settings in
  Facebook with an audience view}. In: Proceedings of the First Conference on
  Usability, Psychology, and Security. pp. 1--8 (2008)

\bibitem{Malandrino2013}
Malandrino, D., Petta, A., Scarano, V., Serra, L., Spinelli, R., Krishnamurthy,
  B.: {Privacy awareness about information leakage: Who knows what about me?}
  In: Proceedings of the 12th ACM Workshop on Privacy in the Electronic
  Society. pp. 279--284 (2013)

\bibitem{Marreiros2016}
Marreiros, H., Tonin, M., Vlassopoulos, M., Schraefel, M.: {`Now that you
  mention it': A survey experiment on information, salience and online
  privacy}. In: CESifo Working Paper, Ludwig Maximilian University of Munich
  (2016)

\bibitem{OxfordEnglishDictionary2016}
{Oxford English Dictionary}: {Salience} (2016),
  \url{http://www.oed.com/view/Entry/170000}

\bibitem{OxfordEnglishDictionary2016a}
{Oxford English Dictionary}: {Taxonomy} (2016),
  \url{http://www.oed.com/view/Entry/198305}

\bibitem{Rajivan2016}
Rajivan, P., Camp, J.: {Influence of privacy attitude and privacy cue framing
  on android app choices}. In: Twelfth Symposium on Usable Privacy and Security
  (2016)

\bibitem{Reamer1979}
Reamer, F.: {Protecting research subjects and unintended consequences: The
  effect of guarantees of confidentiality}. Public Opinion Quarterly  43(4),
  497--506 (1979)

\bibitem{Schaub2015}
Schaub, F., Balebako, R., Durity, A., Cranor, L.F.: {A design space for
  effective privacy notices}. In: Proceedings of the Eleventh Symposium On
  Usable Privacy and Security. pp. 1--17 (2015)

\bibitem{Schaub2016}
Schaub, F., Marella, A., Kalvani, P., Ur, B., Pan, C., Forney, E., Cranor, L.:
  {Watching them watching me: Browser extensions' impact on user privacy
  awareness and concern}. In: NDSS Workshop on Usable Security (2016)

\bibitem{Stemler2001}
Stemler, S.: {An overview of content analysis}. Practical Assessment, Research
  {\&} Evaluation  7(17),  137--146 (2001)

\bibitem{Stutzman2010}
Stutzman, F., Kramer-Duffield, J.: {Friends only: Examining a privacy-enhancing
  behavior in Facebook}. In: Proceedings of the SIGCHI Conference on Human
  Factors in Computing Systems. pp. 1553--1562 (2010)

\bibitem{Thomas2006}
Thomas, D.: {A general inductive approach for analyzing qualitative evaluation
  data}. American Journal of Evaluation  27(2),  237--246 (2006)

\bibitem{Tsai2009}
Tsai, J., Egelman, S., Cranor, L., Acquisti, A.: {The impact of privacy
  indicators on search engine browsing patterns}. In: Proceedings of the Fifth
  Symposium on Usable Privacy and Security (2009)

\bibitem{Vail2008}
Vail, M., Earp, J., Ant{\'{o}}n, A.: {An empirical study of consumer
  perceptions and comprehension of web site privacy policies}. IEEE
  Transactions on Engineering Management  55(3),  442--454 (2008)

\bibitem{Vemou2014}
Vemou, K., Karyda, M., Kokolakis, S.: {Directions for raising privacy awareness
  in SNS platforms}. In: Proceedings of the 18th Panhellenic Conference on
  Informatics. pp. 1--6 (2014)

\bibitem{Wang2015a}
Wang, N., Zhang, B., Liu, B., Jin, H.: {Investigating effects of control and
  ads awareness on android users' privacy behaviors and perceptions}. In:
  Proceedings of the 17th International Conference on Human-Computer
  Interaction with Mobile Devices and Services. pp. 373--382 (2015)

\bibitem{Wang2014}
Wang, Y., Leon, P.G., Acquisti, A., Cranor, L.F., Forget, A., Sadeh, N.: {A
  field trial of privacy nudges for Facebook}. In: Proceedings of the 32nd
  Annual ACM Conference on Human Factors in Computing Systems. pp. 2367--2376
  (2014)

\bibitem{Williams2016a}
Williams, M., Nurse, J.{\relax R.C}., Creese, S.: {The perfect storm: The
  privacy paradox and the Internet-of-Things}. In: Workshop on Challenges in
  Information Security and Privacy Management at the 11th International
  Conference on Availability‚ Reliability and Security (ARES) (2016)

\bibitem{Wohlin2014}
Wohlin, C.: {Guidelines for snowballing in systematic literature studies and a
  replication in software engineering}. In: Proceedings of the 18th
  International Conference on Evaluation and Assessment in Software
  Engineering. pp. 38--48 (2014)

\bibitem{Yang2015}
Yang, R., Ng, Y., Vishwanath, A.: {Do social media privacy policies matter?
  Evaluating the effects of familiarity and privacy seals on cognitive
  processing}. In: Proceedings of the 48th Hawaii International Conference on
  System Sciences. pp. 3463--3472 (2015)

\bibitem{Ziegeldorf2015}
Ziegeldorf, J.H., Henze, M., Hummen, R., Wehrle, K.: {Comparison-based privacy:
  Nudging privacy in social media (position paper)}. Data Privacy Management,
  and Security Assurance in Lecture Notes in Computer Science  9481,  226--234
  (2015)

\end{thebibliography}

\end{document}